\documentclass[12pt]{iopart}
\usepackage{graphicx}

\begin{document}

\title{Real-time monitoring of complex moduli from micro-rheology}
\author{Taiki Yanagishima $^1$,  Daan Frenkel $^2$, Jurij Kotar $^{1,3}$, Erika Eiser $^1$}
\address{$^1$ Cavendish Laboratory, University of Cambridge, Cambridge,U.K. \\ $^2$ Department of Chemistry, University of Cambridge, Cambridge,U.K. \\ $^3$ Nanoscience Centre, University of Cambridge, Cambridge,U.K.}
\date{}

\begin{abstract}
We describe an approach to online analysis of micro-rheology data using a multi-scale time-correlation method. The method is particularly suited to process high-volume data streams and compress the relevant information in real time. Using this, we can obtain complex moduli of visco-elastic media without suffering from the high-frequency artefacts that are associated with the truncation errors in the most widely used versions of micro-rheology. Moreover, the present approach obviates the need to choose the time interval for data acquisition beforehand. We test our approach first on an artificial data set and then on experimental data obtained both for an optically trapped colloidal probe in water and a similar probe in poly-ethylene glycol solutions at various concentrations. In all cases, we obtain good agreement with the bulk rheology data in the region of overlap. We compare our method with the conventional Kramers-Kronig transform approach and find that the two methods agree over most of the frequency regime. For the same data set, the present approach is  superior to Kramers-Kronig at high frequencies and can be made to perform at least comparable  at low frequencies. 
\end{abstract}


\maketitle

\section{Introduction}
Passive micro-rheology is a powerful experimental technique to probe the viscoelastic properties of liquids \cite{Schnurr1997,Mason1997,Mason1995(2)}. The method can be used to perform {\em in situ} measurements of the linear visco-elastic properties of complex liquids on  samples  that are many orders of magnitude smaller than those required in conventional rheology experiments. In a typical micro-rheology experiment, inert colloids are dispersed throughout the medium under study. The random (Brownian) displacements of the colloids are tracked optically, e.g. by video camera ($\sim$ kHz sampling rate) or using a quadrant photodiode  ($\sim$ MHz sampling rate). In what follows, we focus on ``real space'' micro-rheology experiments, although it should be stressed that the key ideas behind the approach were first put forward by Mason and Weitz~\cite{Mason1995} in the context of diffusing-wave spectroscopy experiments. 

The starting point for all micro-rheology studies is the generalized Stokes-Einstein relation
\begin{equation}\label{eq:GSE}
D(s)=\frac{kT}{6\pi\eta(s)Rs}
\end{equation}
where $D(s)$ is the Laplace transform of the time-dependent diffusion coefficient of a spherical particle with radius $R$ and $\eta(s)$ is the Laplace transform of the time-dependent viscosity, which is related to the Laplace-transformed modulus $G(s)$ by $G(s)=s\eta(s)$. Eqn.~\ref{eq:GSE} was proposed by Mason and Weitz~\cite{Mason1995} in the context of diffusing-wave spectroscopy and implemented for real-space (particle-tracking) micro-rheology by Gittes and Schnurr et al.~\cite{Gittes1997,Schnurr1997}. The approach was placed on a firm theoretical footing by Levine and Lubensky~\cite{Levine2000}.  Excellent reviews of microrheology can be found in refs.~\cite{Mason2000,Gardel2005,Waigh2005,Cicuta2007}. 

In what follows, we assume that the conditions for the validity of Eqn.~\ref{eq:GSE} are satisfied. As micro-rheology is by now a well-established and widely used technique, the focus of the present paper is not on the basic equations of micro-rheology, but rather on the most convenient way to translate experimental data on particle displacements into frequency-dependent, complex moduli. Again, several approaches to convert time series of particle displacements into complex moduli have been discussed in the literature (see, e.g. \cite{Gittes1997,Mason1997,Buchanan2005,Crocker2000a}).  However, the existing approaches suffer from one or more of the following drawbacks: 1) truncation errors in the transformation of the data for time series to frequency dependent moduli, 2) the need to fix the size of the data set (and hence the duration of the measurement) in advance, 3) the use of analytical approximations to the experimental data to facilitate the calculation of the visco-elastic moduli. 

The approach that we propose here has the advantage that it is fast and simple and it does not suffer from truncation errors. Analytical approximations are not needed to obtain an accurate  high-frequency response. Statistical noise may affect the performance at low frequencies. In that case, the best results are achieved by using a properly normalized analytical approximation to the raw data.  Moreover, due to the use of an on-line data-reduction procedure, there is in practice no limit on the size of the data stream (and hence on the duration of the measurement). As a result,  the emerging results for the complex moduli can be viewed ``on line'' during data acquisition. 

Moreover, the data-reduction procedure is effectively ``loss free'' -- that means that one and the same data set can be used to probe the visco-elastic moduli at high and low frequencies. The choice of the low-frequency cut-off can be made during the experiments: by simply running longer, the low-frequency cut-off is decreased. However, we stress that the real advantage of the present method is not at low frequencies, but at high frequencies. This is significant because one of the key advantages of micro-rheology is precisely that it can access the visco-elastic moduli at frequencies that are too high to be probed in conventional rheology experiments.  

The basic idea behind our approach is to compute (during data acquisition) the time-correlation function of particle displacements and use a systematic coarse-graining procedure~\cite{Frenkel2002} that reduces the memory requirements for computing a correlation function from the full set of $N$ points to ${\mathcal O}(\ln N)$ points. Importantly, the method still makes use of all available data points and hence no physically meaningful information is lost in the coarse-graining process.

\begin{figure}[htb]
	  \centering
		\includegraphics[width = 0.7\linewidth]{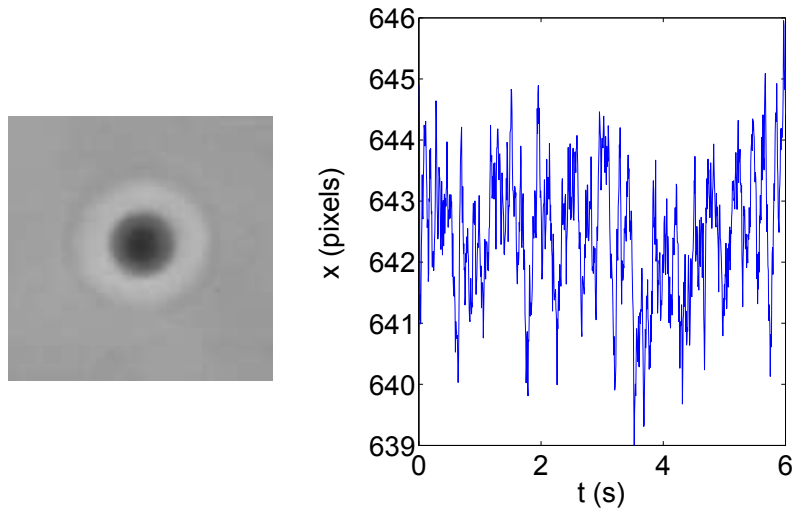}
	\caption{Video camera capture of a $1\mu$m diameter polystyrene microsphere in aqueous solution in an infrared (1064 nm) laser trap, and an example of tracking by image correlation. By tracking the Brownian motion of the particle, one can calculate the viscoelastic properties of the medium.}
	\label{fig:Brownian}
\end{figure}

Below, we describe our approach and show how, in combination with a variable-interval Fourier-Laplace transform method described in ref.~\cite{Evans2009}, it can be used to compute complex moduli that agree with the results obtained by existing techniques (at least, in the region where the latter do not suffer from truncation errors).  We first test our approach on ``synthetic'' data and then we apply it to experimental micro-rheology data.

\section{Computation of complex moduli}

\subsection{Relationship between time series and complex moduli}
We start from Eqn.~\ref{eq:GSE} where the Laplace transform of the diffusion coefficient is given by 
 \begin{equation}\label{eq:defDs}
D(s)\equiv\int_0^\infty dt\; e^{-st} D(t)
\end{equation}
and 
\begin{equation}\label{eq:GK}
D(t)=\int_0^t dt'\;  \langle v_x(0)v_x(t') \rangle
\end{equation}
In Eqn~\ref{eq:GK} we have expressed the time-dependent diffusion coefficient as the integral of the ``velocity autocorrelation function'' (VACF) ${\langle}v_x(0)v_x(t){\rangle}$. Note that the time scales in micro-rheology are such that one never measures the true velocity of the colloidal particle, rather one measures the (diffusive) displacement $\Delta x(\tau)$ of such a particle in the shortest time interval $\tau$. In what follows, we interpret the ``velocity'' $v_x(t)$ as $(x(t+\tau)-x(t))/\tau$. With this definition - and the replacement of the integration in Eqn.~\ref{eq:GK} by a summation, we obtain the expression for the time-dependent diffusion coefficient that we will use below.
Using the generalised Stokes-Einstein expression (Eqn.~\ref{eq:GSE}), we can now relate the Laplace transform of the velocity autocorrelation function to the elastic moduli of the medium. Using eqns.~\ref{eq:defDs} and \ref{eq:GK} above, we can write
\begin{eqnarray}\label{eq:DLaplace}
D(s) & = & \int_0^\infty dt\; e^{-st}  \int_0^t dt' \;\langle v_x(0)v_x(t')\rangle \nonumber \\
& = & \frac{{\mathcal L}(\langle v_x(0)v_x(t)\rangle)}{s} \; ,
\end{eqnarray}
where ${\mathcal L}(f(t))$ denotes the Laplace transform of function $f(t)$. As we shall see below, a similar expression can still be used when we consider a sequence of position measurements at discrete time intervals. Using eqn.~\ref{eq:GSE} we can then write
\begin{equation}
\frac{{\mathcal L}(\langle v_x(0)v_x(t)\rangle)}{s}=\frac{kT}{6\pi R G(s)}
\end{equation}
If the colloidal particle is confined in an optical trap, then the force constant $\kappa$ of this trap  
appears as a correction to the elastic modulus:
\begin{equation}
6\pi R G(s) \rightarrow 6\pi R G(s) + \kappa
\end{equation}
This expression is (of course) also correct in the absence of an embedding medium (i.e. with $G(s)=0$), in which case
\begin{equation}
D(s)=\frac{kT}{\kappa}
\end{equation}
The mean-squared displacement of the particle is then
\begin{equation}
\langle (\Delta x)^2 (t\rightarrow\infty)\rangle=  2\lim_{s\rightarrow 0} s\frac{D(s)}{s}
\end{equation}
or
\begin{equation}
\langle (\Delta x)^2 (t)\rangle=  2\frac{ kT}{\kappa}
\end{equation}
as expected from equi-partition, where ${\mathcal L}^{-1}$ is the inverse Laplace transform.
In the limit $\kappa\rightarrow 0$ and $G(s)=s\eta$ we recover the Einstein relation
\begin{equation}
\langle (\Delta x)^2 (t)\rangle=  2Dt \;.
\end{equation}
In general, our expression for $G(s)$ is
\begin{equation}\label{eq:Gs}
G(s) = \frac{kT}{6\pi R D(s)}-\frac{\kappa}{6\pi R}
\end{equation}
Our key point is that it is advantageous to determine $G(s)$ directly from the velocity auto-correlation function  (or, more precisely, the discrete-time equivalent of this object). This correlation function is then coarse-grained online and $D(s)$ follows directly from a single Fourier Laplace transform.

Note that the conventional method to analyse micro-rheology data starts by computing the power spectrum of the particle displacements $\left\langle  |x(\omega)|^2\right\rangle$. This power spectrum is related to imaginary part of the complex susceptibility $\alpha(\omega)$, that is related to the complex modulus $G^*(\omega)$ through
\begin{equation}
G^*(\omega) = \frac{1}{6{\pi}R\alpha^*(\omega)}
\end{equation}
where $^*$ denotes a complex quantity.
The imaginary part of  $\alpha(\omega)$ is related to $\left\langle  |x(\omega)|^2\right\rangle$ through
\begin{equation}
\alpha''(\omega) = \frac{\omega}{4k_BT}\left\langle |x(\omega)|^2 \right\rangle
\end{equation}
In order to obtain $G(\omega)$, we need to know both $\alpha'$ and $\alpha''$. 
$\alpha'$ can be obtained from $\alpha''$ via a Kramers-Kronig transform
\begin{eqnarray}\label{eq:KKalpha} 
\alpha' (\omega) & =  & \frac{2}{\pi}P\int_0^{\infty}\frac{\zeta\alpha''(\zeta)}{\zeta^2 - \omega^2}d\zeta \nonumber\\
& = & \frac{2}{\pi}\int_0^{\infty}\cos(\omega t) dt \int^{\infty}_0 \alpha''(\zeta)\sin(\zeta t)d(\zeta)
\end{eqnarray}
Note that the Kramers-Kroning route requires several transforms instead of one. Moreover, as is obvious from the integration limits of eqn.~\ref{eq:KKalpha}, the KK transform is sensitive to the high-frequency cut-off of the data. This tends to manifest itself as apparent unphysical behavior of the computed moduli at the limits of the frequency interval studied. Moreover, we shall show that the method based on the velocity correlation function allows a very convenient ``on-line'' reduction of the data, such that with very moderate storage (${\mathcal O}(\log N)$ -- where $N$ is the number of data points) -- we retain the full information about the complex moduli over a frequency interval that extends from $2\pi/T$ to $2\pi/\Delta t$, where $T$ is the total time of the measurement and $\Delta t$ is the interval between successive data points $n+1$ and $n$. 

Below, we briefly describe how we perform this on-line coarse-graining of the incoming data stream. The method used is similar to the approach used to compute time correlation functions over very long time intervals in computer simulations~\cite{Frenkel2002} and for on-board data reduction in Dynamic Light Scattering experiments \cite{Peters1993}.

\subsection{Coarse-graining procedure}
Raw data, in the form of particle displacements, are continuously fed into a circular buffer of length $N$ (e.g. $N=100$) in a cyclic fashion, such that once it is filled, the first data point (say point 1) is overwritten by the most recent data (say point N+1). The primary data are of the form $\delta x(n\Delta t)\equiv x((n+1)\Delta t)-x(n\Delta t)$, where $x$ is the (calibrated) displacement of the particle that is being tracked and $\Delta t$ the time interval between successive sampling points. The index $n$ runs from 1 to $n_{max}$, where $n_{max}$ is the total number of points sampled during the measurement. 

To achieve coarse graining, we sum the $M$ most recently entered data points whenever $n\pmod{M}=0$ and enter the result into the next level 'coarse-grained' array that is also organised as a circular buffer. In this procedure, the size of the arrays for each level of coarse graining is preset and, as the number of coarse-graining levels grows only logarithmically with the total number of data points, overfilling the arrays is, in practice, impossible. The only important requirement is that the computation of the correlation function (see below) is faster than the data acquisition rate. In the example that we study, this is always the case. But even for data-acquisition speeds in the MHz regime, on-line processing of the data should not be a problem (for instance, by using a Graphics Processing Unit).  If, for instance, we consider the case with $N$=100 points per array, and a coarse graining factor $M$=10, then the total storage required to accumulate the correlation function for a run of a billion points would be $2\times 8\times 100$, including the ''accumulator arrays'' (see below).

\begin{figure}[htb]
	\centering
		\includegraphics[width = 0.7\linewidth]{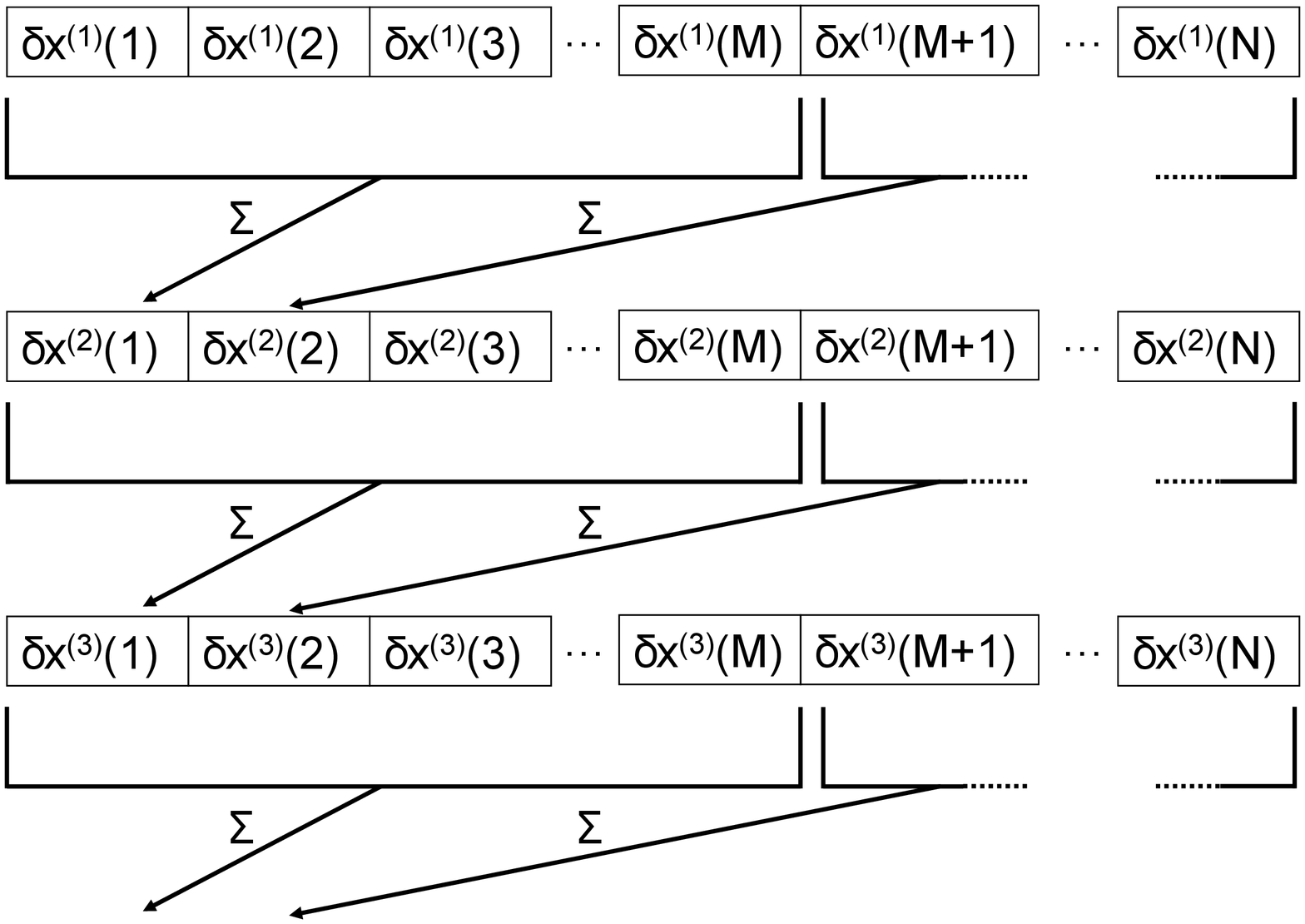}
	\caption{An example of data flow in a coarse graining scheme \cite{Frenkel2002} using	an $N$-point arrays and a coarse-graining factor of $N/M$. New data is cyclically introduced into the top (Level 1) array $\delta x^{(1)}(t_i)$, and averaged into Level 2 'coarse-grained' arrays with longer times $\delta x^{(2)}(t'_i)$. These are subsequently averaged into an even higher Level 3 $\delta x^{(3)}(t_i)$ etc.}
	\label{fig:CG}
\end{figure}

Data in the successive coarse-graining buffers are related such that
\begin{equation}
\delta x^{(i+1)}(t)=\sum_{n=1}^M \delta x^{(i)}(n\Delta t)\; ,
\end{equation}
i.e.  $M$ data points from level $i$ are summed and entered into the appropriate bin of the ${i+1}^{\textrm{th}}$ buffer.

In addition to the circular data buffers, we use an equal number of ``accumulator'' arrays where we store the information about the  velocity auto-correlation function. Whenever a new data point is entered into an array of size $N$ at coarse-graining level $m$, it is  multiplied with itself and with all $N - 1$ preceding points in the same buffer.  The resulting $N$ products are accumulated in a linear array of length $N$, such that the product of points $N$ and $N-n$ are added to bin $n+1$. For every level of coarse graining there is a separate accumulator array. Computing the $N$ products and adding the resulting numbers to the appropriate elements of the accumulator array is the most time-consuming part of our data acquisition algorithm. However, even with MHz acquisition rates, modern processors (in particular GPU's) should easily be able to keep up with the data acquisition.

The key assumption in the procedure described here is that the correlation functions that are measured show no high-frequency modulations at long times. This would clearly be incorrect if the particles would, for instance, perform a high-frequency, undamped periodic motion. Yet, for Brownian motion in a passive medium, this assumption is always satisfied. 

At the end of an experiment, we have an estimate of the desired ``velocity'' auto-correlation function. However, it is stored in different arrays -- one for each level of coarse graining.  Moreover, these arrays overlap. This is important because this allows us to check if the coarse graining introduces systematic errors at short times; for the first point, it always does. However, as we will show below, the effect of coarse-graining is minimal for subsequent points. 

The final step in our analysis is to construct a single correlation function out of the various coarse-grained arrays and apply a Fourier-Laplace transform to compute $D(s)$. Each subsequent level of the correlation function has a time spacing that is $M$ times larger than that at the previous level and extends for over a correspondingly longer time. By combining different levels of coarse graining, we are thus extending the low frequency limit of our micro-rheology data, while the high frequency limit is still set by the Nyquist condition.

\begin{figure}[htb]
	\label{VACF}
	\centering
	\includegraphics[width = 0.7\linewidth]{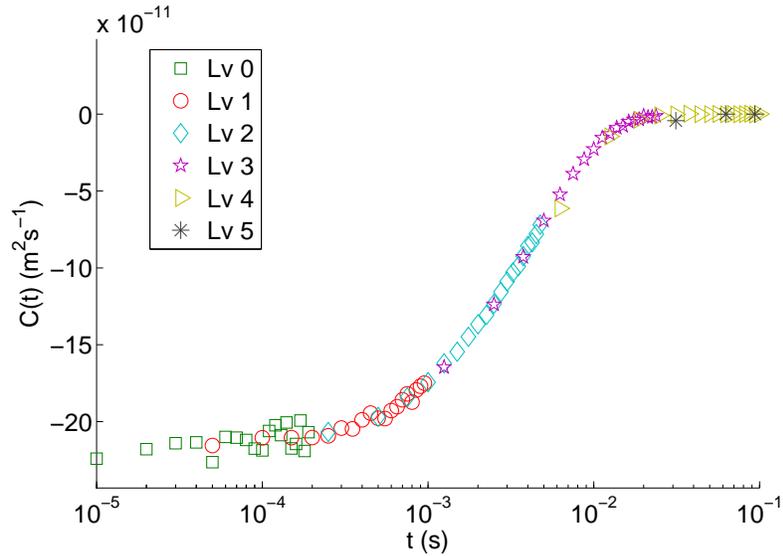}
	\caption{An example of a series of coarse-grained correlation functions, in this case VACFs calculated from simulated fluctuations of a 1$\mu$m sphere in an optical trap with trapping constant $\kappa = 10^{-3}$ pN/nm, diffusing with diffusion constant $D = 10^{-12}$ m$^2$s$^{-1}$. Correlations found from different levels are found to overlap.}
\end{figure}

\subsection{Direct conversion of coarse-grained time series into complex moduli}
In order to obtain $G^*(\omega)$,we first need to compute the Fourier-Laplace transform of the VACF. Due to our coarse-graining procedure, we have values for the VACF at non-equidistant points in time. A convenient procedure to obtain the Laplace transform of such a data set was proposed by Evans et al. \cite{Evans2009}. The method uses a linear interpolation to connect successive data point. Between data points, the slope of these linear segments is (of course) constant, but at every data point, the slope may change discontinuously. The second derivative of this interpolated function is zero everywhere, except at the data points where the slope of the linear interpolation changes. There, the second derivative is a delta-function with an amplitude equal to the change in slope of the linear segments. 

\begin{figure}[htb]
  \centering
	\includegraphics[width = 0.7\linewidth]{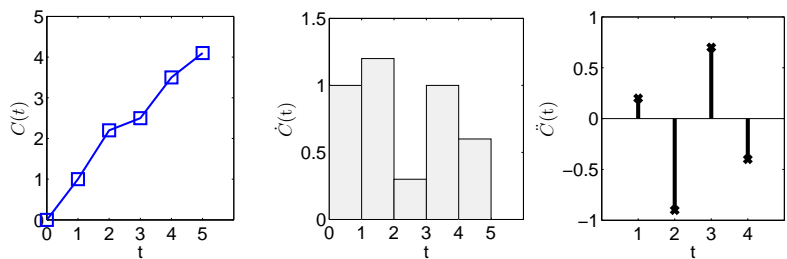}
	\label{deltadiagram}
	\caption{Transformation of an experimentally obtained correlation is found by considering a linear piecewise function which passes through all points and considering the functional form and magnitude of the first and second derivative. If $C(t)$ is the original correlation function, $\dot{C}(t)$ and $\ddot{C}(t)$ are the first and second derivatives respectively.}
\end{figure}

\begin{eqnarray}
\ddot C(t_i) & = & \left(\frac{C(t_{i+1}) - C(t_i)}{t_{i+1}-t_i} - \frac{C(t_{i}) - C(t_{i-1})}{t_{i}-t_{i-1}}\right)\delta(t-t_i) \nonumber \\
& \equiv & a_i \delta(t-t_i)
\end{eqnarray}
where $a_i$ is a shorthand notation of the amplitude of the delta function at $t=t_i$. The Laplace transform of a series of delta functions is 
\[
{\mathcal L}(\ddot C)=\sum_i a_i e^{-st_i}
\]
for all $t_i\ne 0$.

Next, consider the contribution of the point at $t=0$. It is important to note that this first point of the VACF has an amplitude that is much larger than any of the points at $t\ne 0$. For this reason, it is convenient to treat this point -- and its Laplace transform -- separately from that of the rest of the VACF. To this end, we split the VACF into a delta-function part at $t=0$ and a remainder that has been defined such that it is continuous, and has a continuous first derivative, at $t=0$. This is achieved as follows: we define $C_\delta(t)$, the delta-function part at $t=0$ as
\begin{equation}\label{eq:Cdelta}
C_\delta(t=0)=\left(C(t=0) - C(t=\Delta t)\right)\Delta t\delta(t)
\end{equation}
where $C(t=\Delta t))$ is the value of the VACF after one time step.
The remainder of the correlation function is the same as before, except that the value at $t=0$ is equal to the value at $\Delta t$. As a consequence, this function is continuous and has a continuous first derivative at $t=0$. Hence, the second derivative of the linear interpolation  of this function vanishes at $t=0$. We first compute the Laplace transform of the part delta-function part.
It is given by 
\[
\frac{\left(C(t=0) - C(t=\Delta t)\right)}{2}\Delta t
\]
The factor $1/2$ follows from the fact that the delta function is symmetric around $t=0$ when obtained as a limit of a delta function -- we only integrate the part with $t>0$.  Note that integration of this part of the VACF yields a constant, real contribution to the diffusion coefficient, $D_{c}$. For confined particles  $D(t\rightarrow\infty)=0$. However, due to statistical noise, this condition may not be satisfied. In that case, it can be enforced  by adding a small correction to the first point of the VACF to ensure that $D(t\rightarrow\infty)=0$:
\[
C(t=0) \rightarrow C(t=0) - 2 \times \frac{D_c}{\Delta t} \;.
\]
Next, we consider $C_{VV}(s)$, the Laplace transform of the remainder of the VACF.
We use that fact that
\[
{\mathcal L}(\ddot f) = s^2f(s)-sF(0)-\dot F(0)
\]
Hence,
\begin{equation}\label{eq:LaplCvv}
C_{VV}(s) = \frac{{\mathcal L}(\ddot C)}{s^2} +\frac{C_{VV}(t=0)}{s}+0
\end{equation}
where the final zero on the right hand side follows because the first derivative of $C_{VV}(t)$ vanishes at $t=0$. It then follows that
\begin{equation}\label{eq:Cseries}
C_{VV}(s) = \frac{\sum_i a_i e^{-st_i}}{s^2} +\frac{C_{VV}(t=0)}{s}
\end{equation}
This part of the Laplace transform of the VACF must be added to the constant contribution that results from the transform of the delta function at $t=0$.
Combining \ref{eq:Cseries} and \ref{eq:Cdelta}, we get a complete Laplace transform of the VACF $C(s)$, which is related to the diffusion constant $D(s)$ using \ref{eq:DLaplace} and thus the complex moduli via \ref{eq:Gs}. In what follows, we shall consider the Fourier-Laplace transform, i.e. we replace $s$ by $-\imath\omega$.

\section{Test of method}
\begin{figure}[htb]
	\centering
		\includegraphics[width = 0.7\linewidth]{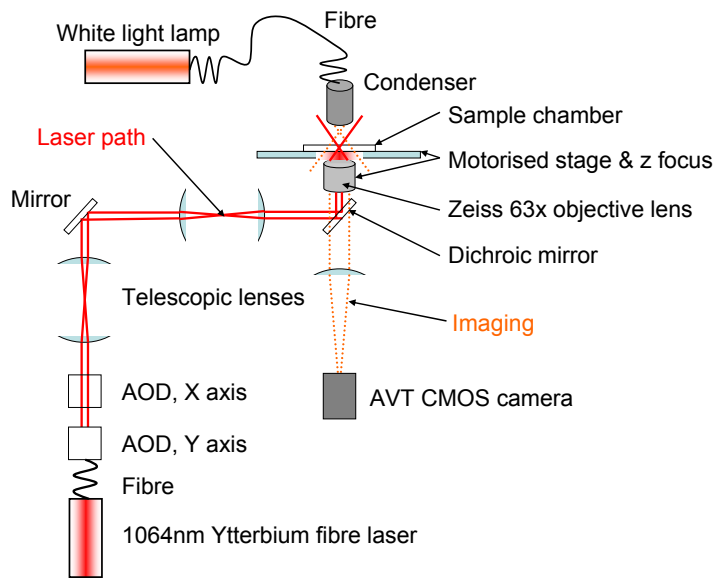}
	\caption{The optical tweezer setup involved the use of a 1064 nm infrared laser, an AOD (acouto-optic deflector) unit and three stepping motors for moving the sample in the xy and z planes respectively. Visualisation was obtained using a CMOS camera.}
	\label{fig:tweezer}
\end{figure}
We tested the approach described above both on real and on``synthetic'' micro-rheology data. The synthetic data were obtained numerically, using a kinetic Monte Carlo simulation of a particle in a harmonic trap. The objective of this test was to verify that the diffusion coefficient and the trap stiffness that were used as input parameters in the simulation, would be recovered in the subsequent data analysis of the VACF.

Experimental data were obtained using an infrared optical tweezer setup that is described in ref.~\cite{Leoni2009}. The laser beam was focused by a 63x water immersion objective with numerical aperture 0.90. The sample chamber consisted of a standard microscope slide, 100 $\mu$m SecureSeal imaging spacers (Grace Biolabs) and a coverslip. The chamber was inverted to ensure that the laser light was introduced from the cover-slip side. The sample was moved in the xy plane using a motorised stage, and an acousto-optic deflector (AOD) unit was used to accurately position the trap. Movement of the trap in the z direction was achieved using a separate motor moving the lens vertically.

The motion of microspheres was captured using a CMOS camera (Marlin F-131B, Allied Vision Technologies) with a maximum frame rate of around 500 frames per second. This maximum rate was only achieved with the field of view restricted to one bead. To track particles, we used spatial correlation with an optimised kernel followed by a 2 dimensional fit. Pixels were converted into distances using a graticule.

The samples investigated were dilute (around 1 particle per 10 $\mu m ^3$) suspensions of 3$\mu$m diameter monodisperse silica micro-spheres (Microspheres GmbH) suspended in solutions of polyethylene glycol (PEG 8000) in MilliQ water at PEG concentrations of 0, 20 and 40 \% by weight.

For the sake of comparison, we used bulk rheology measurements (Physica MCR benchtop rheometer (Anton Paar GmbH) ) to obtain independent estimates of $G'$ and $G''$ for these solutions. The viscosity data thus obtained were in good agreement with the literature data reported in ref.~\cite{Gonzalez-tello2002}.

We used MATLAB for the data analysis and the Monte Carlo simulations.

\section{Results and discussion}
\begin{figure}[htb]
\centering
\includegraphics[width = 0.7\linewidth]{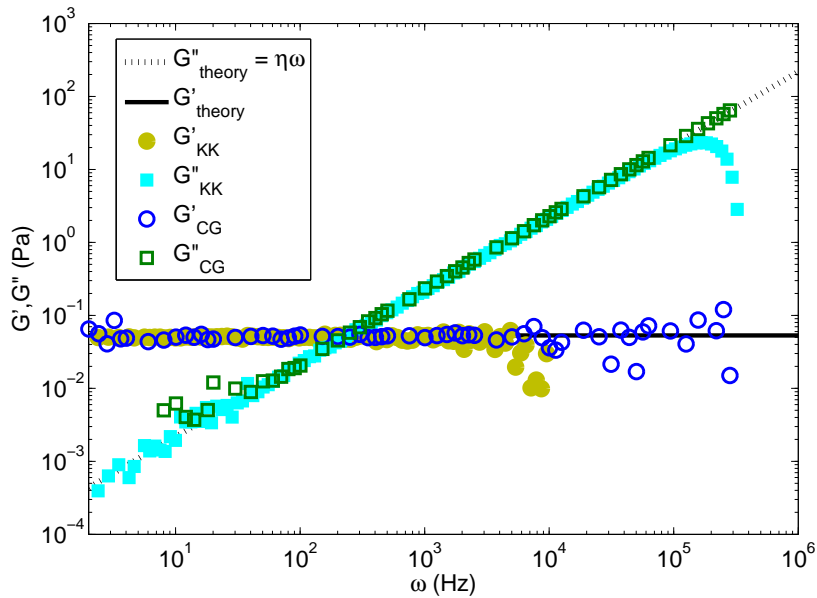}	
	\caption{Synthetic data for a Brownian particle moving in a harmonic trap. The elastic response of the trap results in an effective elastic modulus $G'$.  In the same figure, we also show the results of the conventional analysis, using the Kramers-Kronig method. We note that the KK method suffers from systematic truncation errors at high frequencies. In contrast, the present method suffers from statistical noise at low frequencies. These problems can be addressed by replacing the (noisy) long-time part of the VACF by an analytical fit. See Figure \ref{fig:FitGstar}}.
	\label{fig:FreeTrapped}
	\end{figure}
\begin{figure}[htb]
\centering
\includegraphics[width = 0.7\linewidth]{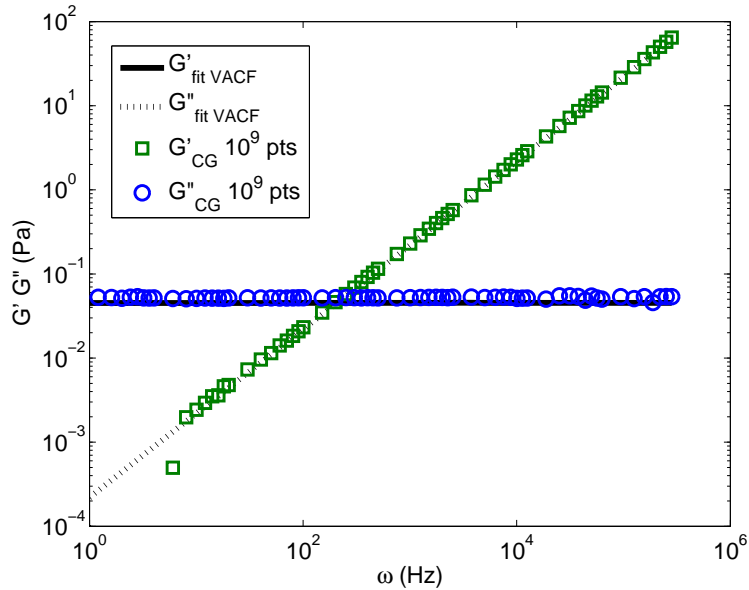}	
	\caption{Analysis of ``synthetic'' data for the same set of parameters as in Fig.~\ref{fig:FreeTrapped} but with a data set that was 100 times larger (10$^9$ points).  The better statistics result in a significant reduction of the noise in  the estimated visco-elastic moduli at both high and low frequencies. Note that, as in Fig.~\ref{fig:FreeTrapped}, G' is the ``apparent'' elastic modulus that is due to the stiffness of the harmonic trap. The drawn and dotted curves were obtained by fitting the VACF to a delta-function plus a single exponential. Clearly, such a  fitting procedure further suppresses the effect of the statistical noise.}
	\label{fig:FitGstar}
	\end{figure}
We start with the analysis of the ``synthetic'' data. We generated an artificial trajectory of a Brownian particle in a harmonic trap, using the following parameters: diffusion constant $D = 10^{-12}$ m$^{2}$s$^{-1}$, temperature $T = 298$K and sampling rate 100 kHz. An artificial trap wsa imposed with $\kappa = 10^{-3}$ pN/nm stiffness. As the harmonic trap was isotropic, moduli obtained from the X and Y displacements could be averaged. In most of our studies, the length of the data sets  was $10^7$ points (100s). To relate the diffusion constants to visco-elastic moduli, we chose a value of  $1{\mu}$m for  the radius $R$ of the microsphere. Using the Stokes-Einstein relation, a diffusion coefficient $D = 10^{-12}$ m$^{2}$s$^{-1}$ then corresponds to a viscosity $\eta = 2.18 \times 10^{-4}$ Pas. The synthetic data correspond to a particle in a purely viscous liquid. In the case of a particle in an harmonic trap, the apparent elastic modulus $G'$ follows from Eqn.~\ref{eq:Gs}.  

The moduli found are shown in Figure \ref{fig:FreeTrapped}. In this figure, we have not corrected for the trap stiffness to obtain the true $G'$  because we compare the effective $G'$ with the value that follows from the known spring constant of the trap. The (apparent) real modulus is indeed found to agree with the predicted value $G'_{trap} = 0.0531$ Pa. $G"$ follows the expected $\eta\omega$ curve.

To illustrate the difference between the present approach and the KK method, we have plotted the moduli as obtained by the Kramers-Kronig transformation method \cite{Gittes1997} in Figure \ref{fig:FreeTrapped}. As can be seen from the figure, the KK approach becomes unreliable at high frequencies. This problem has also been reported by Schnurr et al. \cite{Schnurr1997}. In contrast, the present method appears robust in the frequency region where the KK method fails. This difference is important as the high-frequency sampling of the visco-elastic properties is  one of the main strengths of micro-rheology: the use of a method that is robust at the highest frequencies is therefore important.

\begin{figure}[htb]
	\centering
	\includegraphics[width = 0.7\linewidth]{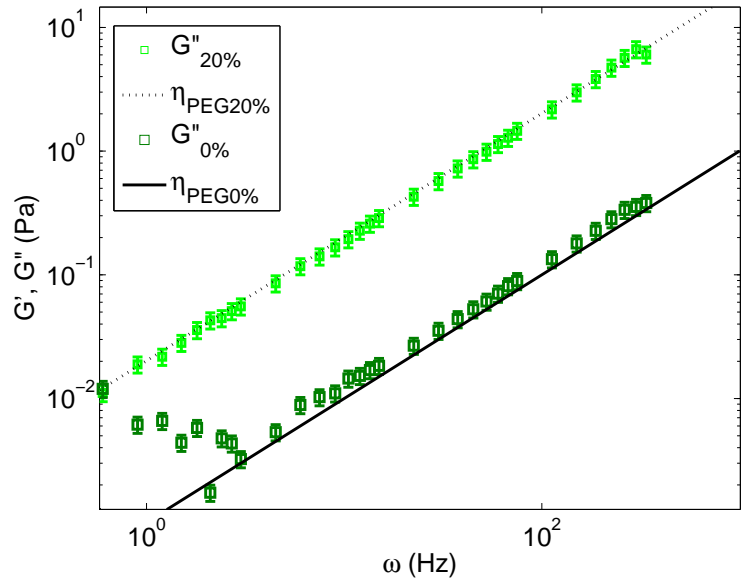}
	\caption{Calculation of G'' from experimental micro-rheology data on pure water and on a 20\% wt PEG8000 solution (see text). In this figure G' is not shown because, after correction for the trap stiffness, it does not differ significantly from zero.  The straight lines are the predictions for G'' based on the known low-frequency viscosities (see  Ref.~\cite{Gonzalez-tello2002}).}
	\label{fig:PEG0020}
\end{figure}
\begin{figure}[htb]
	\centering
		\includegraphics[width = 0.7\linewidth]{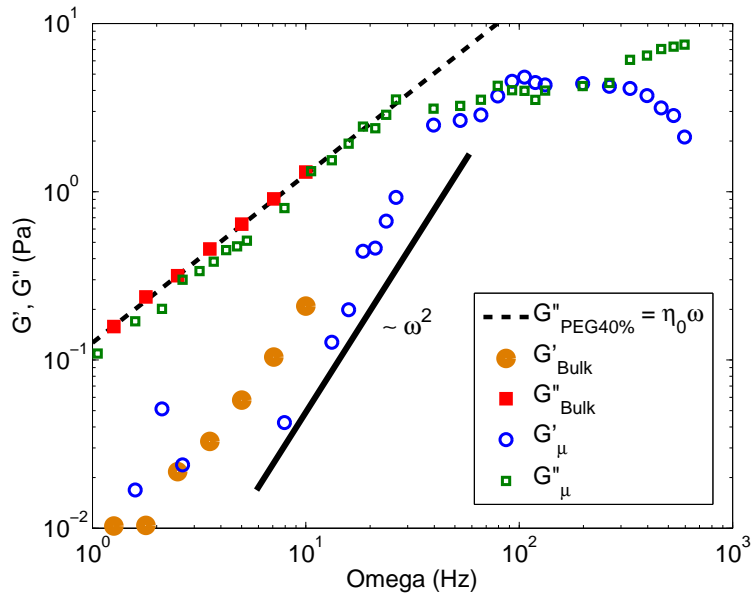}
	\caption{Comparison of the complex moduli G' and G" of a  40\% wt aqueous PEG8000 solution as obtained by classical rheology (closed symbols) and micro-rheology (open symbols). As this solution exhibits visco-eleastic behaviour, G' shows the (expected) quadratic increase with frequency, whereas G"  agrees well with the zero-frequency viscosity at low frequencies (126.5 mPa s~\cite{Gonzalez-tello2002}) and levels off at near the point where G' peaks. As can be seen, the micro-rheology results are largely complementary to the conventional viscosity data. In regions of overlap, the two methods agree. Note that the micro-rheology data become very noisy for moduli less that 5 10$^{-2}$ Pa. Longer runs (and/or fitting the VACF) would alleviate this problem.}
		\label{fig:PEG40}
\end{figure}At low frequencies, where $G''\ll G'$, the visco-elastic moduli become noisy. The reason is that, whereas in the KK approach, a positive sign of $G''$ is guaranteed even when $G''\ll G'$, this constraint is not imposed for the present method. This problem is due to statistical noise and can easily be fixed by fitting the $t>0$ part of the VACF to a sum of exponentials with the constraint that the VACF (i.e. including the contribution at $t=0$) integrates to zero (at least, for particles in a trap).  Figure \ref{fig:FitGstar} shows how the use of such a constrained fit greatly improves the quality of the low-frequency moduli. Data fitting can be avoided altogether by simply collecting a larger data set. Here, we benefit from the fact that the storage required by our method scales only logarithmically with the size of the data set. As an illustration, Fig.~\ref{fig:FitGstar}  shows the results for $G'$ and $G"$ obtained from a set of 1 billion (synthetic) data points; analysing such a data set with the conventional (KK) algorithm would require at least 8 gigabytes of memory.

Data sets were also taken for the suspensions of silica beads in 3 different concentrations of PEG 8000. In all three cases, we used the same strength of laser trapping as before. The bead diameters were 3${\mu}$ for the 0 and 20\% measurements, 1$\mu$m for the 40\% - the moduli should be independent of bead size. The curves shown in Figs.~\ref{fig:PEG0020} and ~\ref{fig:PEG40} are the result of averaging moduli obtained using x and y displacement data. The error bars arise from polydispersity of the beads used.

Interestingly, the viscosity of the most concentrated 40\% wt aqueous PEG solution appears to decrease at high frequencies. This suggests that the stress autocorrelation function (that is related by a Green-Kubo relation to the viscosity) does not decay completely on the shortest time scales sampled in the experiments. This observation may be related to the non-Newtonian behaviour that has been observed in high molecular weight polyethylene oxide \cite{Ebagninin2009}. In the high frequency range, we also observe a corresponding increase in the elastic modulus with an intermediate $\sim\omega^2$ region, characteristic of a Maxwell fluid \cite{Ferry1948}.

At lower frequencies, the micro-rheology experiments yield data that are in good agreement with those that we obtained in bulk rheology measurements. Of course, the high-frequency range cannot be probed with bulk rheology.

\section{Conclusions}

In conclusion, we have developed a coarse-graining scheme that provides an efficient means to obtain complex moduli from micro-rheology measurements in real time. Even for very long measurements, the required memory usage of the method is very small. This is important, as illustrated by  recent work on the high bandwidth measurement of the VACF of a Brownian particle \cite{Huang2010}. In this paper, it is stressed that the direct accumulation of the VACF is limited by the memory capacity of the data acquisition hardware. The approach produced here should present an easily implemented solution to this, and many other experimental studies related to micro-rheology.

\ack{We would like to thank Dr Pietro Cicuta for kindly giving us access to his optical tweezer setup. This work was funded by the following bodies: Ernest Oppenheimer Fund (TY); the George and Lillian Schiff Foundation (TY); the Royal Society of London via the Wolfon Merit Award (DF); the European Research Council (Advanced Grant agreement 227758) (DF); the Cavendish Laboratory, Cambridge, UK (EE) and the BP Institute for multiphase flow (EE).}

\newpage

\bibliographystyle{iopart-num}
\bibliography{biblio}

\end{document}